\documentclass[aps,prc,preprint,amsmath,showpacs,superscriptaddress,nofootinbib]{revtex4-1}
\newcommand{\beqy}{\begin{eqnarray}}
\newcommand{\eeqy}{\end{eqnarray}}
\newcommand{\bmlet}{\begin{subequations}}
\newcommand{\emlet}{\end{subequations}}
\usepackage{graphicx}% Include figure files
\usepackage{dcolumn}% Align table columns on decimal point
\usepackage{bm}% bold math
\usepackage{hyperref}% add hypertext capabilities

\def\apj{ApJ}  % Astrophysical Journal
\def\apjl{ApJ} % Astrophysical Journal, Letters
\def\prl{Phys.~Rev.~Lett.} % Physical Review Letters
\def\pr{Phys.~Rev.} %Physical Review
\def\prc{Phys.~Rev.~C} % Physical Review C
\def\aap{A\&A}   % Astronomy and Astrophysics
\def\mnras{MNRAS}             % Monthly Notices of the RAS
\def\cpc{Chin.~Phys.~C}%European Physical Journal A
\begin{document}

\title{Role of the crust on the tidal deformability of a neutron star within a unified treatment of dense matter}

\author{L. Perot}
\affiliation{Institute of Astronomy and Astrophysics, Universit\'e Libre de Bruxelles, CP 226, Boulevard du Triomphe, B-1050 Brussels, Belgium}
\author{N. Chamel}
\affiliation{Institute of Astronomy and Astrophysics, Universit\'e Libre de Bruxelles, CP 226, Boulevard du Triomphe, B-1050 Brussels, Belgium}
\author{A. Sourie}
\affiliation{Institute of Astronomy and Astrophysics, Universit\'e Libre de Bruxelles, CP 226, Boulevard du Triomphe, B-1050 Brussels, Belgium}
\affiliation{LUTH, Observatoire de Paris, PSL Research University, CNRS, Universit\'e Paris Diderot,  Sorbonne Paris Cit\'e, 5 place Jules Janssen, 92195 Meudon, France.}

\begin{abstract}
The role of the crust on the tidal deformability of a cold nonaccreted neutron star is studied using the recent unified equation of state BSk24. This equation of state, which is based on the nuclear-energy density functional theory, provides a thermodynamically consistent description of all stellar regions. Results obtained with this equation of state are compared to those calculated for a putative neutron star made entirely of homogeneous matter. The presence of the crustal layers is thus found to significantly reduce the Love number $k_2$, especially for low-mass stars. However, this reduction mainly arises from the increase in the stellar radius almost independently of the equation of state. This allows for a simple analytic estimate of $k_2$ for realistic neutron stars using the equation of state of homogeneous matter only. 
\end{abstract}

\maketitle

\section{Introduction}
\label{sec:intro}

The first direct detection of the binary neutron star (NS) merger GW170817 by the LIGO-Virgo collaboration~\cite{ligo2017inspiral} has provided a new way to probe the dense-matter equation of state (EoS) through observations of the tidal deformations of the two  inspiralling NSs (see, e.g., Ref.~\cite{baiotti2019} for a recent review). 

The question arises as to what extent the tidal deformability or polarizability of a NS, as measured by the dimensionless parameter 
\begin{equation}
    \label{eq:Lambda}
    \Lambda=\dfrac{2}{3} \,   k_2 \, \left(\dfrac{c^2 R}{GM}\right)^5
\end{equation} 
(with $R$ the circumferential radius of the star, $M$ its gravitational mass, and $k_2$ the second gravito-electric Love number, $c$ the speed of light, $G$ the gravitational constant), depends on its complex internal structure (see, e.g., Ref.~\cite{blaschke2018} for a recent review about the different phases of dense matter in a NS). Beneath a very thin atmosphere and possibly a surface ocean, the interior of a NS consists of a solid crust on top of a liquid core. The crust can be further divided into two distinct regions: the outer part made of a crystal lattice of bare atomic nuclei embedded in a charge compensating electron gas, and the inner part characterized by the presence of a neutron liquid. 

The importance of the crustal regions for the determination of the Love number $k_2$ and the tidal deformability $\Lambda$ of a NS has been recently investigated~\cite{piekarewicz2019,kalaitzis2019,ji2019}. 
In Ref.~\cite{piekarewicz2019}, the EoS of the outer crust was calculated as in Ref.~\cite{bps1971} using the nuclear mass model of Duflo \& Zuker~\cite{DZ1995}. The EoS of the core and the crust-core transition were determined from a series of different relativistic mean-field models. A polytropic parametrization was adopted to interpolate the EoS between the outer crust and the core. The role of the inner crust was studied by varying the polytropic index. A different approach was followed in Ref.~\cite{kalaitzis2019}, where different tabulated EoSs were fitted using a sum of two polytropes for energy densities higher than $100$~MeV/fm$^3$ corresponding roughly to the crust-core boundary. Results were compared to those obtained with fitted EoSs extrapolated to lower energy densities. The authors of Ref.~\cite{ji2019} studied the role of the crust on the tidal deformability of NSs by matching the same EoS of the core to different crustal EoSs based on different relativistic mean-field models (the outer crust being described in all cases by the EoS of Ref.~\cite{bps1971}); however, all but one models were found to be incompatible with LIGO-Virgo constraint at 90\% confidence level on the tidal deformabilities of the two NSs inferred from the analysis of GW170817. All of these studies concluded that $k_2$ is very sensitive to the crustal EoS unlike the observable tidal deformability parameter $\Lambda$. However, these studies lead to widely different estimates for the contribution of the crust on $\Lambda$, ranging from $0.1$\%~\cite{piekarewicz2019} to about $11$\%~\cite{kalaitzis2019} for a $1.4 M_\odot$ NS. Even though the crust was found to have a rather small impact on the tidal deformability, uncertainties on the crustal EoS may still lead to nonnegligible systematic errors on the NS radii inferred from the analysis of the gravitational-wave signal~\cite{gamba2019}. Additional errors may be incurred by the \textit{ad hoc} matching of different crust and core EoSs~\cite{fortin2016,biswas2019}. 

In this paper, the role of the crust on the tidal deformability of a NS is more closely examined using the recent unified EoS based on the generalized Skyrme nuclear-energy density functional BSk24~\cite{pearson2018}. This EoS, whose main features are briefly summarized in Section~\ref{sec:eos}, provides a thermodynamically consistent description of all regions of a NS, from the surface to the central core of the star. The underlying functional was precision fitted to a large set of experimental and theoretical nuclear data~\cite{gcp2013}. We have recently shown that this unified EoS is also consistent with the various constraints inferred from the analyses of the gravitational and electromagnetic observations of GW170817~\cite{perot2019}. To better assess the importance of the crust, results for the Love number $k_2$ and the tidal deformability parameter $\Lambda$ are compared in Section~\ref{sec:tidal} to those obtained using the EoS of a purely liquid putative NS made entirely of homogeneous matter calculated with the same functional BSk24. 

\section{Consistent equations of state for neutron stars with and without crust}
\label{sec:eos}

The EoS we adopt here~\cite{pearson2018} was calculated in the framework of the nuclear energy-density functional theory with the generalized Skyrme effective interaction BSk24~\cite{gcp2013}. The parameters of this functional were primarily fitted to the 2353 measured masses of atomic nuclei having proton number $Z\geq8$ and neutron number $N\geq8$ from the 2012 Atomic Mass Evaluation (AME)~\cite{AME2012}, with nuclear masses being calculated using the self-consistent Hartree-Fock-Bogoliubov (HFB) method allowing for deformations. This functional provides an equally good fit to the 2408 measured masses of nuclei with $N,Z \geq 8 $ from the 2016 AME~\cite{AME2016} with a root-mean-square deviation of 565 keV, as indicated in Table 2 of Ref.~\cite{pearson2018}. This functional was simultaneously adjusted to the EoS of homogeneous neutron matter as calculated in Ref.~\cite{ls2008} within the Brueckner-Hartree-Fock approach using the Argonne V18 nucleon-nucleon potential and a microscopic three-body force. This microscopic EoS is consistent with more recent calculations based on chiral effective field theory~\cite{lynn2016,drischler2019}. With these features, the functional BSk24 is particularly well-suited for a unified description of the crust and core of a NS. 

\subsection{Unified equation of state with crust}

The EoS of the outer crust for mass-energy densities $\rho \gtrsim 10^6$~g~cm$^{-3}$ was calculated in Ref.~\cite{pearson2018} using the latest experimental nuclear mass data from the 2016 AME~\cite{AME2016} supplemented by the recent measurements of copper isotopes~\cite{welker2017}, as well as predictions from HFB calculations using BSk24 for masses that have not yet been measured~\cite{gcp2013}. The following sequence of equilibrium nuclides was found with increasing depth: $^{56}$Fe, $^{62}$Ni, $^{64}$Ni, $^{66}$Ni, $^{86}$Kr, $^{84}$Se, $^{82}$Ge, $^{80}$Zn, $^{78}$Ni, $^{80}$Ni, $^{124}$Mo, $^{122}$Zr, $^{121}$Y, $^{120}$Sr, $^{122}$Sr, and $^{124}$Sr (the first eight elements being fully determined by experimental data). In the region beneath, at mean baryon number densities exceeding $n_\textrm{nd}=2.56\times 10^{-4}$~fm$^{-3}$ (or equivalently $\rho_\textrm{nd}=4.25\times10^{11}$~g~cm$^{-3}$), the crust is permeated by a neutron ocean. The EoS of the inner part of the crust was obtained from the fourth-order extended Thomas-Fermi method with consistent proton shell and pairing corrections added perturbatively~\cite{pcgd2012,pcpg2015}. Most regions of the inner crust were found to be made of zirconium isotopes. At densities above $n_\textrm{pd}=0.073$~fm$^{-3}$ (or equivalently $\rho_\textrm{pd}=1.23\times 10^{14}$~g~cm$^{-3}$), the neutron sea  becomes significantly enriched with free protons. The crust-core transition was predicted to occur at the density $n_\textrm{cc}=0.081$~fm$^{-3}$ (or equivalently $\rho_\textrm{cc}=1.36\times 10^{14}$~g~cm$^{-3}$). Full results can be found in Ref.~\cite{pearson2018}. The EoS over the entire range of mass-energy densities $\rho$ relevant for NSs was conveniently fitted with the following analytic function with relative errors below 1\% in the crustal region of interest here (the errors amount to about 4\% at most in the core)~\cite{pearson2018}:
\begin{align}
\label{fit_function}
\log_{10} (P-P_0) =&\,\frac{p_1+p_2\xi+p_3\xi^3}{1+p_4\xi}\biggl\{\exp[p_5(\xi-p_6)]+1\biggr\}^{-1} +(p_7+p_8\xi)\biggl\{\exp[p_9(p_6-\xi)]+1\biggr\}^{-1} \nonumber\\ &+(p_{10}+p_{11}\xi)\biggl\{\exp[p_{12}(p_{13}-\xi)]+1\biggr\}^{-1}+(p_{14}+p_{15}\xi)\biggl\{\exp[p_{16}(p_{17}-\xi)]+1\biggr\}^{-1} \nonumber\\ &+\frac{p_{18}}{1+[p_{20}(\xi-p_{19})]^2}+\frac{p_{21}}{1+[p_{23}(\xi-p_{22})]^2}\, ,
\end{align}
where $P$ is the pressure and $\xi=\log_{10}\rho$. The term $P_0$ 
accounts for the outer envelope at densities $\rho<10^6$~g~cm$^{-3}$, where atoms are not completely ionized and thermal effects cannot be neglected. According to Ref.~\cite{potekhin2013}, the best interpolation to the OPAL EoS~\cite{rogers1996} at $T=10^7$~K  for a partially ionized iron envelop with $P_0=A\,\rho$ yields $A=3.5\times10^{14}$~(cm/s)$^2$. 

\subsection{Unified equation of state without crust}
\label{sec:unified_without}
\begin{figure}[ht!]
\begin{center}
\includegraphics[width=\textwidth]{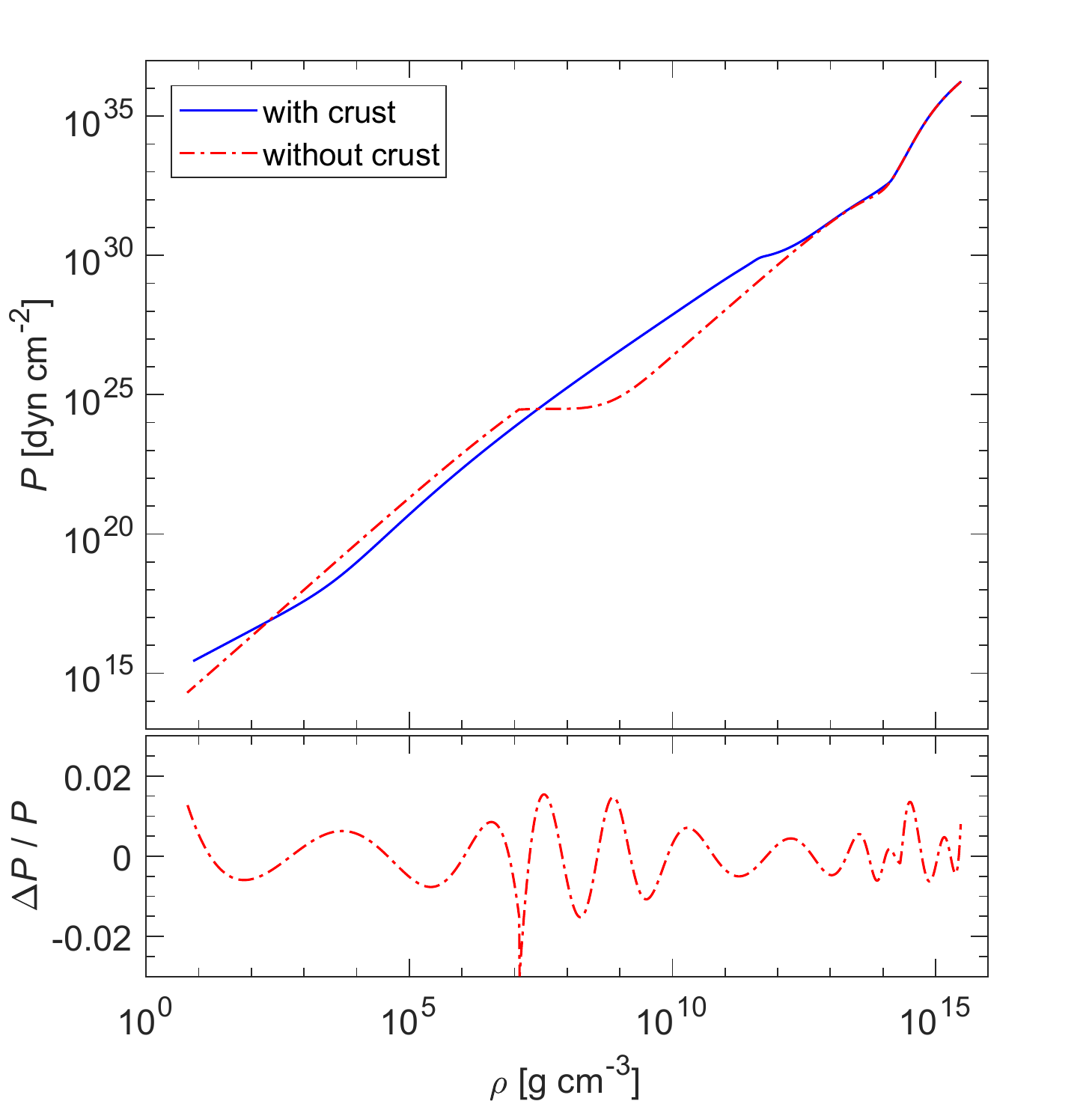}
\caption{(Color online) Top panel: fits of the pressure $P$ as a function of mass-energy density $\rho$ for the unified EoS (`with crust') and the EoS of homogeneous matter (`without crust'). Bottom panel: relative error of the fit with respect to the data, $(P_{\rm fit}-P_{\rm data})/P_{\rm data}$, for the EoS without crust.}
\label{fig:P-RHO-ERROR}
\end{center}
\end{figure}

To investigate the role of the crustal layers on the global structure and on the tidal deformability of a NS, we have extended the EoS of the homogeneous core to densities below $n_\textrm{cc}$ using the same functional BSk24. Of course, such an EoS is unrealistic since homogeneous matter is unstable against nuclear clustering at such densities; this EoS is constructed for the sake of comparison only. The composition is determined from the requirement of electric charge neutrality and from the $\beta$ equilibrium condition
\begin{equation}\label{eq:beta-eq}
    \mu_n=\mu_e+\mu_p\, ,
\end{equation}
where $\mu_i$ denotes the chemical potential of the species $i$ (including the rest-mass energy), whose expressions can be found in Ref.~\cite{pearson2018}. At very low densities, neutrons are unstable and disintegrate by $\beta$ decay into protons, electrons and electron antineutrinos (the latter escaping from the star). The threshold density, as determined by the condition~(\ref{eq:beta-eq}) with $\mu_n=m_n c^2$ ($m_n$ denoting the neutron mass), is found to be $n_\beta=7.33\times 10^{-9}$ fm$^{-3}$ (or equivalently $\rho_\beta=1.23\times 10^7$~g~cm$^{-3}$). The resulting EoS, fitted to the same analytical function~(\ref{fit_function}) (dropping $P_0$) with similar errors, is compared to the EoS with the crust in Fig.~\ref{fig:P-RHO-ERROR}. The fitting parameters $p_i$ are indicated in Table~\ref{fit_table}, for the pressure $P$ in units of dyn~cm$^{-2}$ and the density $\rho$ in g~cm$^{-3}$. The average deviation between the exact and fitted results for $P$ is $1.6\times10^{-3}$\%, reaching $-3.0$\% at the density $\rho_{\beta}$. 

\begin{table}[h!]	
\begin{center}	
\begin{tabular}{c c c}
\hline $i$ &~~~~~~& $p_i$\\
\hline
1&&21.8531\\ 
2&&-0.491085\\
3&&-0.002271\\
4&&-0.037759\\
5&&1.74421\\ 
6&&9.11175\\
7&&13.7367\\
8&&1.15318\\
9&&1.79316\\
10&&-296.748\\
11&&17.8427\\
12&&1.28257\\
13&&15.4300\\
14&&-9.00908\\
15&&1.27191\\
16&&-9020.00\\
17&&7.09244\\
18&&-0.756623\\
19&&14.0815\\
20&&-1.97833\\
21&&15.5396\\
22&&15.2898\\
23&&-0.603784\\
\hline
\end{tabular}	
\end{center}
\caption{\label{fit_table} Fitting parameters for the equation of state without crust, as given by Eq.~(\ref{fit_function}) for the pressure $P$ in units of dyn~cm$^{-2}$ and the density $\rho$ in g~cm$^{-3}$.}
\end{table}	

\subsection{Adiabatic index}

\begin{figure}[ht]
\begin{center}
\includegraphics[width=\textwidth]{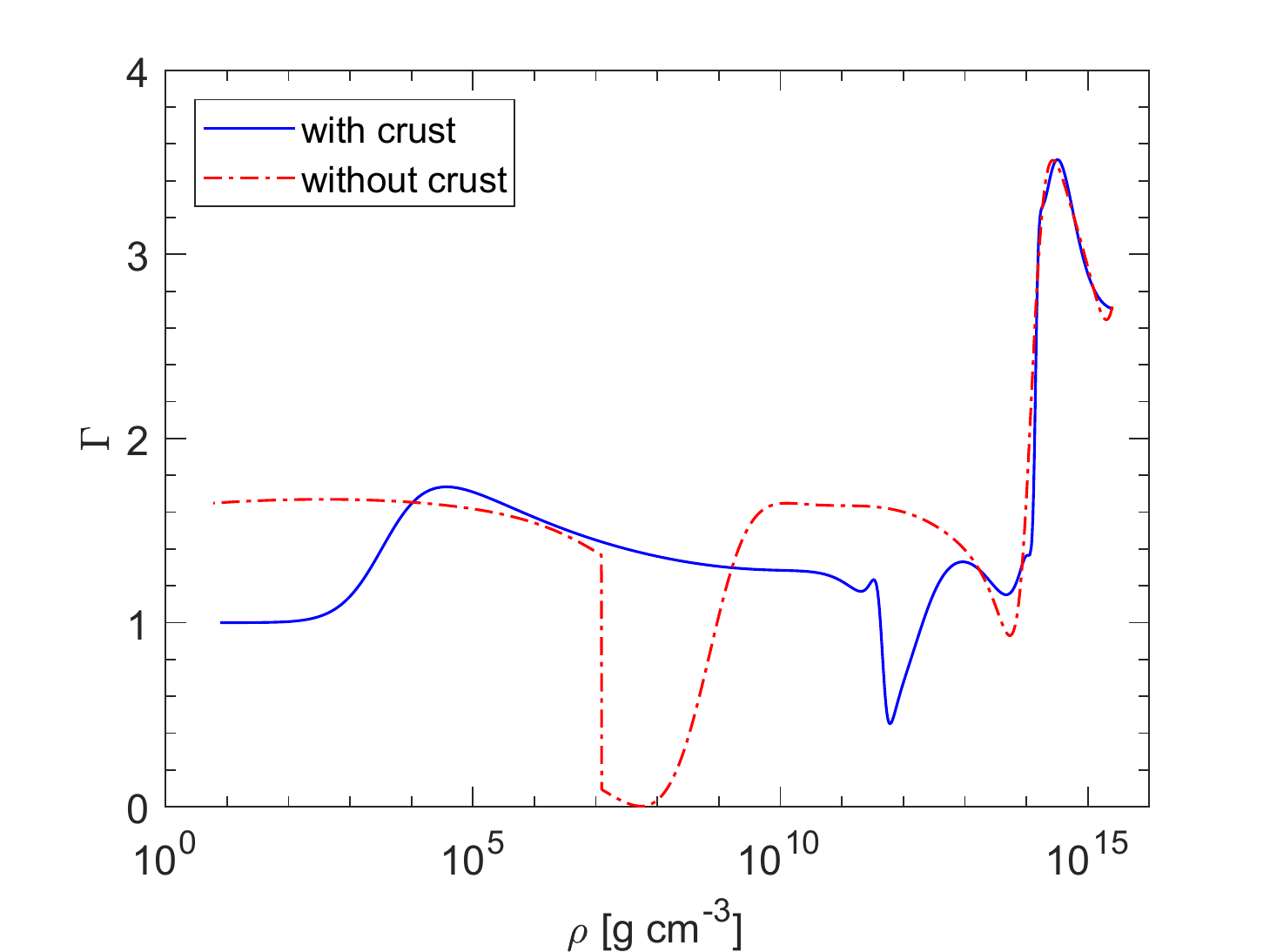}
\caption{(Color online) Variations of the adiabatic index $\Gamma$ with respect to the mass-energy density $\rho$
for the unified EoS (`with crust') and the EoS of homogeneous matter (`without crust').
}
\label{fig:GAMMA-RHO}
\end{center}
\end{figure}

As shown in Fig.~\ref{fig:GAMMA-RHO}, the adiabatic index defined by
\begin{equation}
    \Gamma=\frac{n}{P}\frac{dP}{dn}=\left(1+\frac{P}{\rho\, c^2}\right)\frac{\rho}{P}\frac{dP}{d\rho} \, ,
\end{equation}
varies widely throughout the crust region and can hardly be accurately parametrized by a single polytrope. In the surface envelope, the pressure is proportional to the density and the adiabatic index is therefore equal to $\Gamma=1$. With increasing density, the pressure is mainly determined by that of the degenerate electron Fermi gas. The adiabatic index thus increases to $\Gamma=5/3$ at densities $\rho\sim 10^4-10^5$~g~cm$^{-3}$, and decreases to $\Gamma=4/3$ as electrons become ultrarelativistic. The strong softening of the EoS above $\rho_\textrm{nd}$ marks the transition to the inner crust: the contribution of free neutrons to the pressure is negligible while the mass-energy density is substantially increased. The adiabatic index, which is approximately given by $\Gamma\approx 4/3$ for 
$\rho < \rho_\textrm{nd}$, drops to much lower values for $\rho\gtrsim \rho_\textrm{nd}$, and the difference increases with density as $\delta \Gamma \approx (\rho-\rho_\textrm{nd})^{1/2}$~\cite{bbp1971}. With further compression, the neutron pressure becomes progressively higher and considerably stiffens the EoS with $\Gamma$ approaching $\Gamma \lesssim 2$ at the crust-core boundary. The adiabatic index increases sharply in the outer core due to strong nuclear interactions. The increasingly larger proton fraction at high densities (see Fig. 29 in Ref.\cite{pearson2018}) leads to a softening of the EoS in the central core of the star. 

The EoS of homogeneous matter exhibits very different adiabatic indices. This stems from the fact that both electrons and nucleons contribute to the pressure at all densities. At densities below about $10^4$~g~cm$^{-3}$, electrons and protons form ideal nonrelativistic degenerate Fermi gases therefore  $\Gamma\approx 5/3$. As the degree of relativity increases with density, the adiabatic index declines. The appearance of neutrons at density $\rho_\beta=1.23\times 10^7$~g~cm$^{-3}$ leads to a dramatic softening of the EoS, with the adiabatic index falling to $\Gamma=0$. The subsequent rise of the adiabatic index comes from the increasingly important neutron contribution to the pressure. The nonmonotonic variations of $\Gamma(\rho)$ at higher densities are due to nuclear interactions.

\section{Love number and tidal deformability}
\label{sec:tidal}

The structure of a NS and the Love number $k_2$ are calculated as in our previous work~\cite{perot2019} using the BSk24 EoSs with and without crust (i.e., homogeneous matter throughout the star), as discussed in Sec.~\ref{sec:eos}. In both cases, the hydrostatic equations are integrated from the stellar center to the surface defined by the radial coordinate $R$ where the density $\rho$ has dropped to a few grams per cubic centimeters. Although the fitted EoSs we employ here are continuous, the original tabulated EoS of Ref.~\cite{pearson2018} exhibits density discontinuities at the interface between adjacent layers in the outer crust (density discontinuities are vanishingly small in the inner crust). In principle, density discontinuities require a special treatment (see, e.g., Refs.~\cite{damour2009,postnikov2010,han2019}).  However, the discontinuities are so small in the present case that their impact on $k_2$ and on the global structure of a NS can be safely ignored (the treatment of density discontinuities was originally discussed in the context of compact stars with a quark core assuming that the transition from hadronic to quark matter is very sharp). The global fitting of the EoSs with and without crust leads to slight deviations in the core region, as visible for the highest densities shown in Fig.~\ref{fig:GAMMA-RHO}. However, we have checked that the associated errors on the parameters of the star are negligible: the relative errors on $R$, $k_2$ and $\Lambda$ for a $1.4M_{\odot}$ NS when comparing results using the global fit with parameters in Table~\ref{fit_table} and the same fit for densities below $\rho_{cc}$ combined with the original fit of Ref.~\cite{pearson2018} for the core amount to 0.021\%, -0.53\% and -0.42\%, respectively.

\begin{table}[t]	
\begin{center}	
\begin{tabular}{c c c c c c c c c}
\hline 
$M$ [$M_{\odot}$]&~~~~~~&$r_{\rm cc}$ [km] &~~~~~~&$r_{\rm nd}$ [km]&~~~~~~&$R$ [km]&~~~~~~&$R_0$ [km]\\
\hline
1.4 && 11.54 && 12.11 && 12.59 && 12.13 \\
1.0 && 10.91 && 11.74 && 12.47 && 11.79 \\
\hline
\end{tabular}	
\end{center}
\caption{\label{radius_table} Radial coordinates $r_{\rm cc}$ and $r_{\rm nd}$ at the boundaries between the crust and the core, and the inner and outer regions of the crust respectively; $R$ and $R_0$ denote the circumferential radii of NS having a gravitational mass $M$, with and without crust respectively.}
\end{table}	

\begin{figure}[ht]
\begin{center}
\includegraphics[width=\textwidth]{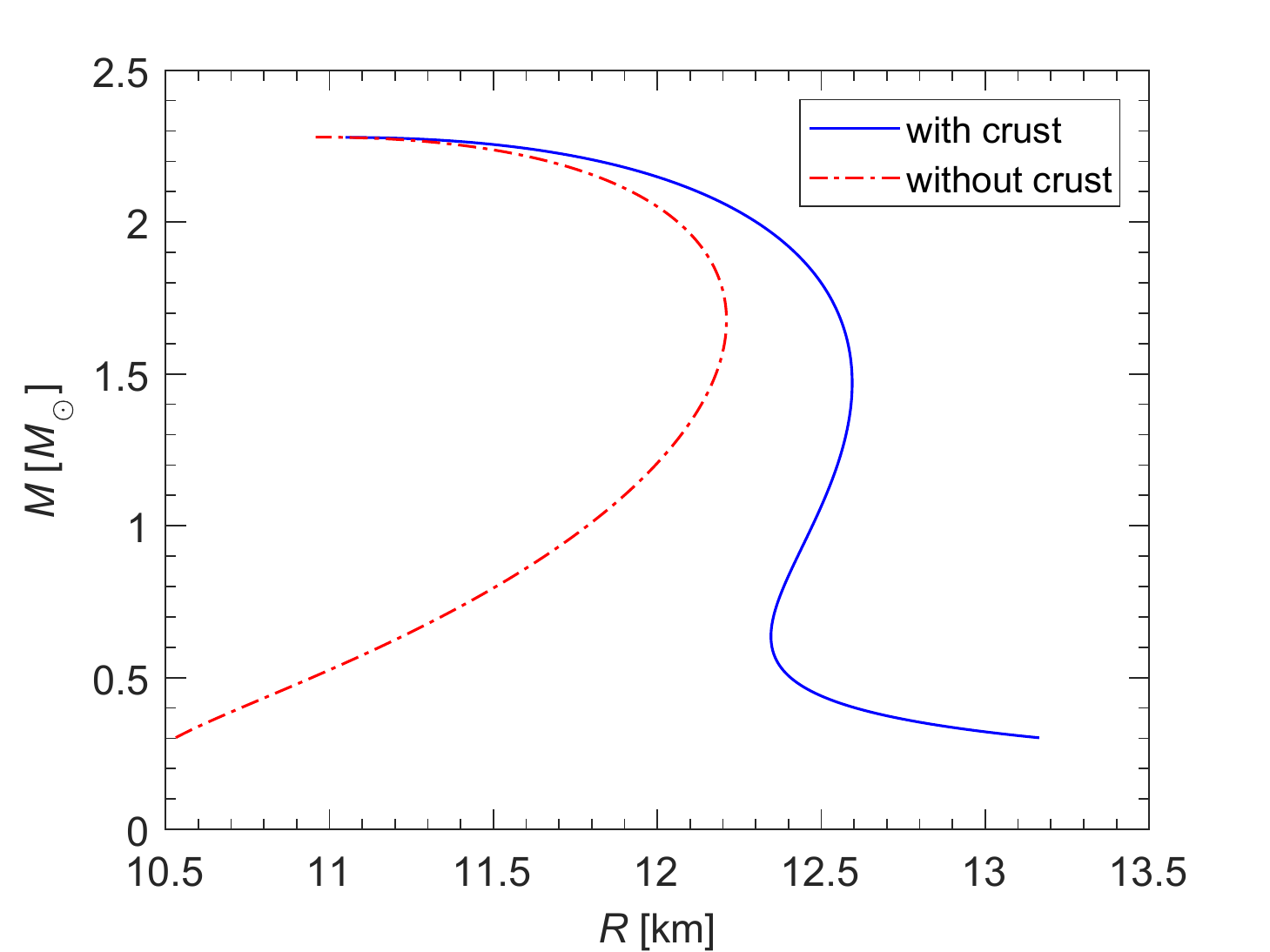}
\caption{(Color online) Gravitational mass $M$ as a function of the circumferential radius $R$ of a nonrotating NS with and without crust.}
\label{fig:M-R}
\end{center}
\end{figure}

\begin{figure}[ht]
\begin{center}
\includegraphics[width=\textwidth]{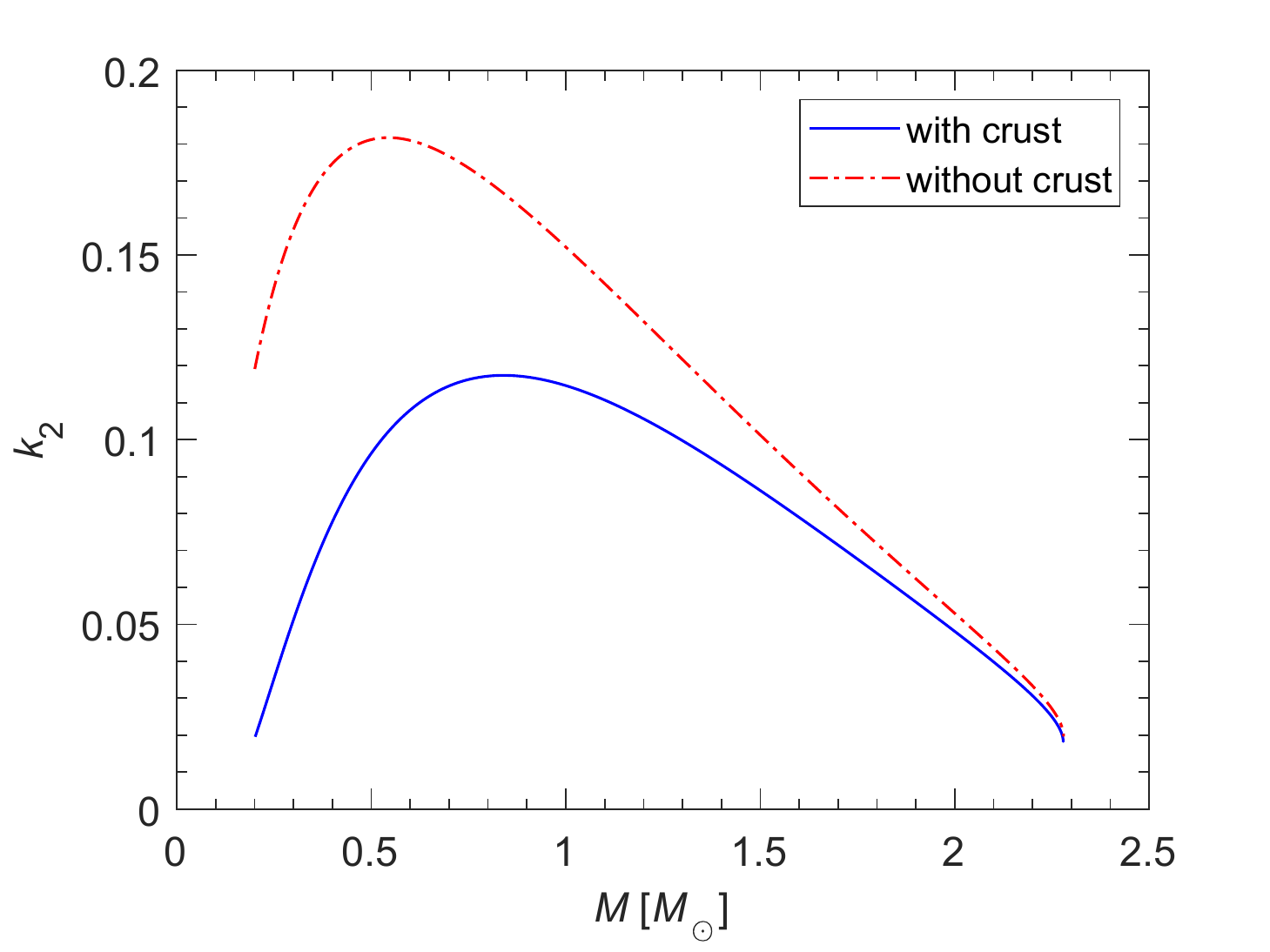}
\caption{(Color online) Second gravito-electric Love number $k_2$ as a function of the gravitational mass $M$ for a NS with and without crust. See text for details.}
\label{fig:k2}
\end{center}
\end{figure}

\begin{figure}[ht]
\begin{center}
\includegraphics[width=\textwidth]{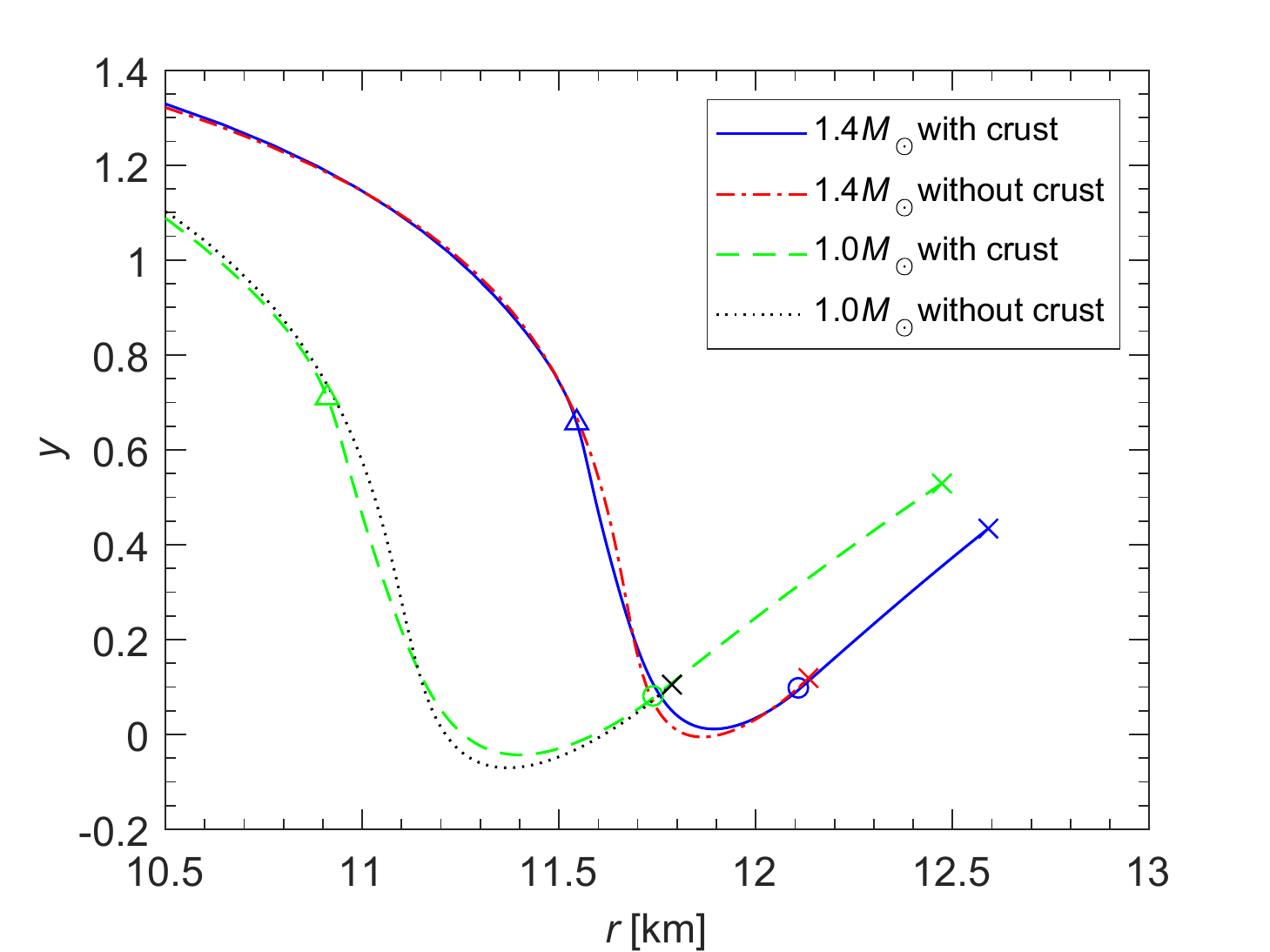}
\caption{(Color online) Function $y(r)$ as a function of the radial coordinate $r$ for NSs with and without crust. The crust-core boundary and the neutron-drip transition are indicated by triangles and circles, respectively. The surface of the star is represented by crosses. See text for details.}
\label{fig:y(r)}
\end{center}
\end{figure}

As shown in Fig.~\ref{fig:M-R}, a NS would have a smaller radius if it were made entirely of homogeneous matter throughout its interior. The radial coordinates $r_\textrm{cc}\equiv r(n_\textrm{cc})$ at the crust-core boundary and $r_\textrm{nd}\equiv r( n_\textrm{nd})$ at the neutron-drip transition (delimiting the boundary between the outer and inner crusts) are indicated in Table~\ref{radius_table}. The less massive the NS is, the more important is the reduction of its radius. However, the differences remain small for realistic neutron-star masses, 
being negligible for the maximum-mass configuration and reaching about 6\% for a one solar mass NS. On the contrary, the crust has a much more dramatic impact on the Love number $k_2$, as shown in Fig.~\ref{fig:k2}. The Love number $k_2$ is given by~\cite{hinderer08,hinderer10} 
\begin{align}
k_2 =& \,\frac{8C^5}{5}(1-2C)^2\big[2+2C(y_R-1)-y_R\big] \, \Big\{2C\big[6-3y_R+3C(5y_R-8)\big] \nonumber \\ &+ 4C^3\big[13-11y_R+C(3y_R-2)+2C^2(1+y_R)\big] \nonumber \\ &+ 3(1-2C)^2\big[2-y_R+2C(y_R-1)\big] \ln (1-2C) \Big\}^{-1} \, ,
\label{eq:k2}
\end{align}
where we have introduced the compactness parameter
\begin{equation}
    C = \dfrac{G\, M}{R\, c^2}\, ,
\end{equation}
and the quantity $y_R\equiv y(R)$ is obtained by integrating the following differential equation~\cite{postnikov2010} (indicating by a prime derivation with respect to $r$) 
\begin{equation}
\label{y_eq}
r y'(r)+ y(r)^2 + F(r) y(r) + Q(r)=0 \, ,
\end{equation}
\begin{equation}
\label{F(r)}
    F(r)=\frac{1-4\pi G r^2(\mathcal{E}(r)-P(r))/ c^4}{1-2Gm(r)/(r c^2)}\, ,
\end{equation}
\begin{align}
\label{Q(r)}
    Q(r)=&\frac{4 \pi G r^2/c^4}{1-2Gm(r)/(r c^2)}\Biggl[5\mathcal{E}(r)+9P(r)+\frac{\mathcal{E}(r)+P(r)}{c_s(r)^2} c^2-\frac{6\,  c^4}{4\pi r^2 G}\Biggr]\nonumber \\
    &-4\Biggl[ \frac{G(m(r)/(r c^2)+4\pi r^2 P(r)/c^4)}{1-2Gm(r)/(r c^2)}\Biggr]^2\, ,
\end{align}
 where $\mathcal{E}(r)= \rho(r) c^2$ denotes the energy density at the radial (circumferential) coordinate $r$, $P(r)$ the pressure, $c_s=c\sqrt{dP/d\mathcal{E}}$ 
 the sound speed, and $m(r)$ is the gravitational mass function obtained from the solution of the Tolman-Oppenheimer-Volkoff equations~\cite{tolman1939,oppenheimer1939}. Equation~(\ref{y_eq}) must be solved with the boundary condition $y(0)=2$. As shown in Fig.~\ref{fig:y(r)}, the functions $y(r)$ with and without crust are remarkably similar. Therefore, the discrepancies in the Love numbers must originate from the differences in the stellar radii, as shown in the next section.  
 
 \subsection{Approximate analytic solution}
 
\begin{figure}[h!]
\begin{center}
\includegraphics[width=\textwidth]{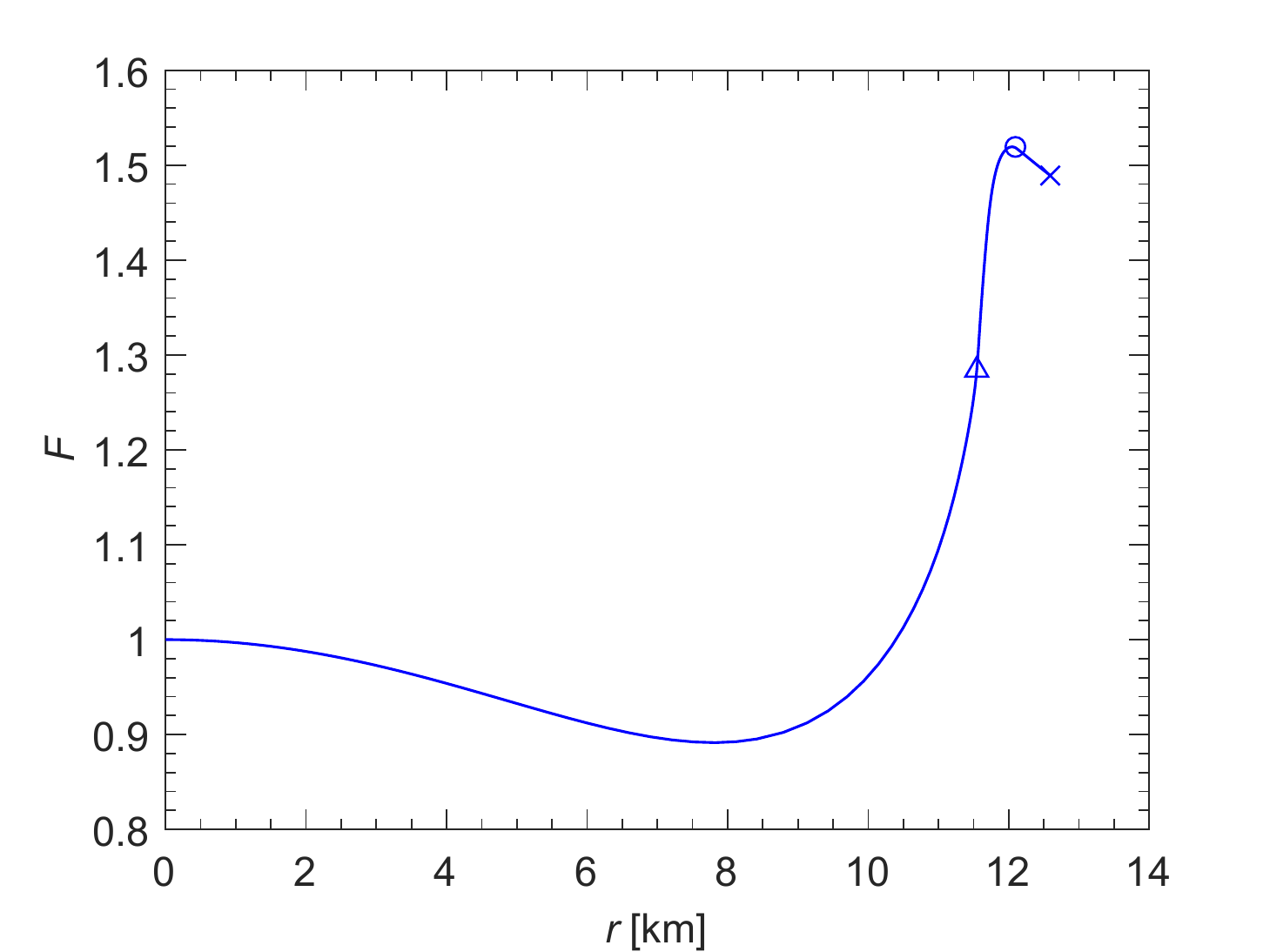}
\caption{Variation of the function $F(r)$ defined by Eq.~(\ref{F(r)}) with the radial coordinate $r$ for a NS with crust having a mass $M=1.4M_{\odot}$. The crust-core boundary and the neutron-drip transition (delimiting the inner and outer parts of the crust) are indicated by a triangle and a circle, respectively. The surface of the star is represented by a cross.}
\label{fig:F(r)}
\end{center}
\end{figure}

\begin{figure}[h!]
\begin{center}
\includegraphics[width=\textwidth]{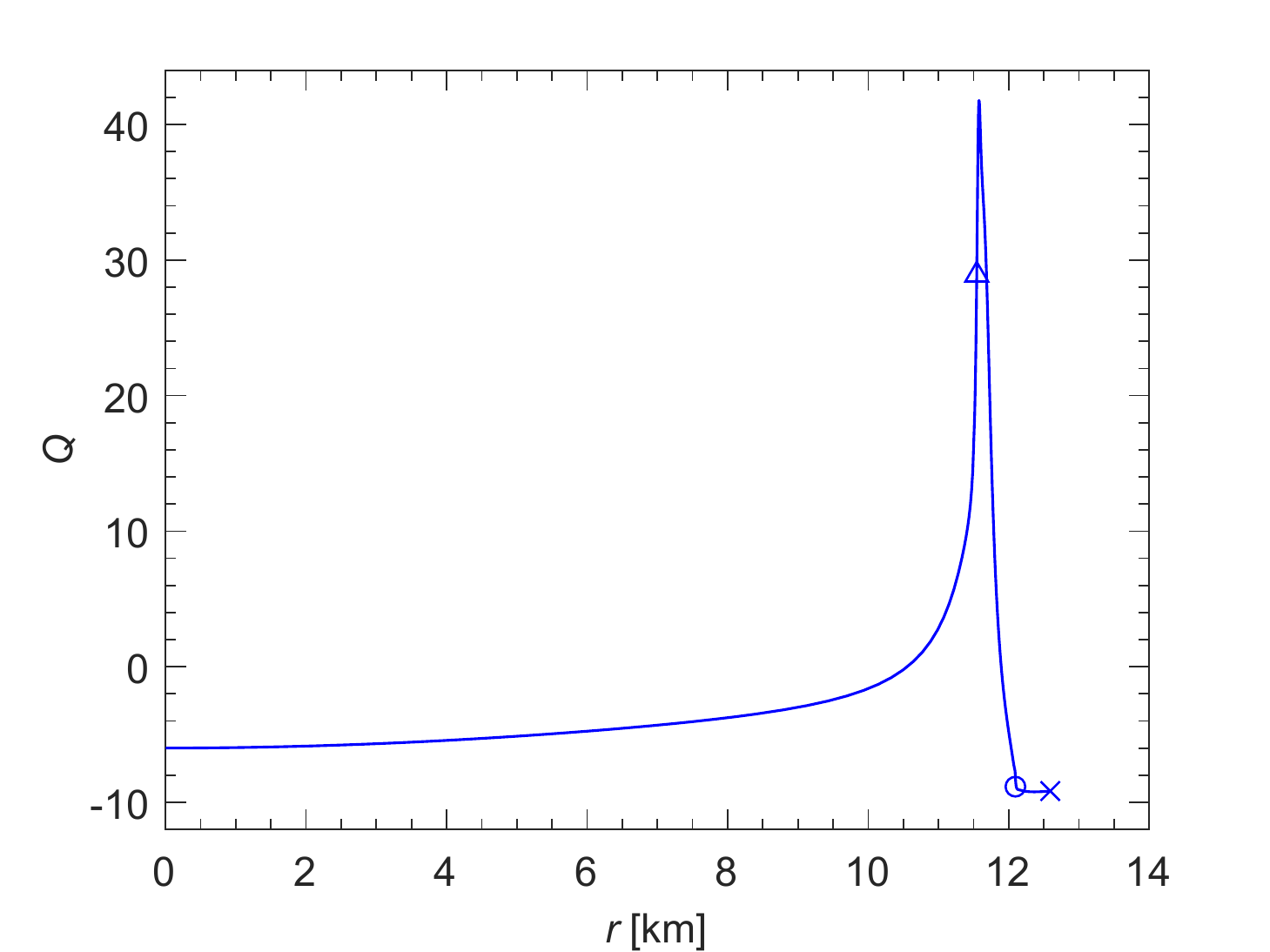}
\caption{Variation of the function $Q(r)$ defined by Eq.~(\ref{Q(r)}) with the radial coordinate $r$ for a NS with crust having a mass $M=1.4M_{\odot}$. The crust-core boundary and the neutron-drip transition (delimiting the inner and outer parts of the crust) are indicated by a triangle and a circle, respectively. The surface of the star is represented by a cross.}
\label{fig:Q(r)}
\end{center}
\end{figure}

\begin{figure}[ht]
\begin{center}
\includegraphics[width=\textwidth]{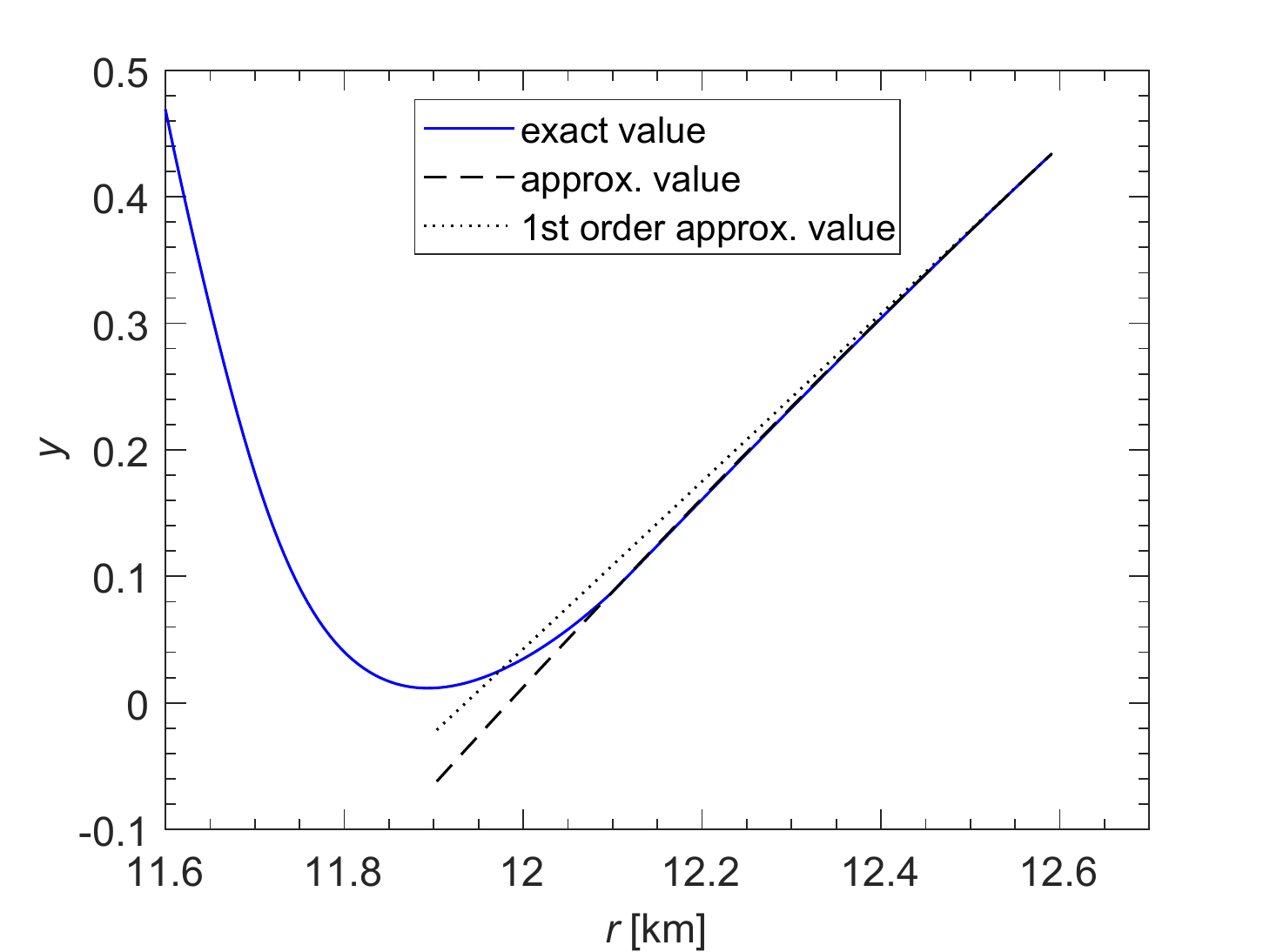}
\caption{(Color online) Variation of the function $y(r)$ with the radial coordinate $r$ for a NS with a mass $M=1.4M_\odot$. Exact values were calculated by solving numerically Eq.~(\ref{y_eq}) using the EoS with crust. Approximate values were obtained using Eq.~(\ref{y_eq_app}) and the first-order expansion  (\ref{y_eq_app1}).}
\label{fig:Y_approx_CRUST}
\end{center}
\end{figure}

\begin{figure}[ht]
\begin{center}
\includegraphics[width=\textwidth]{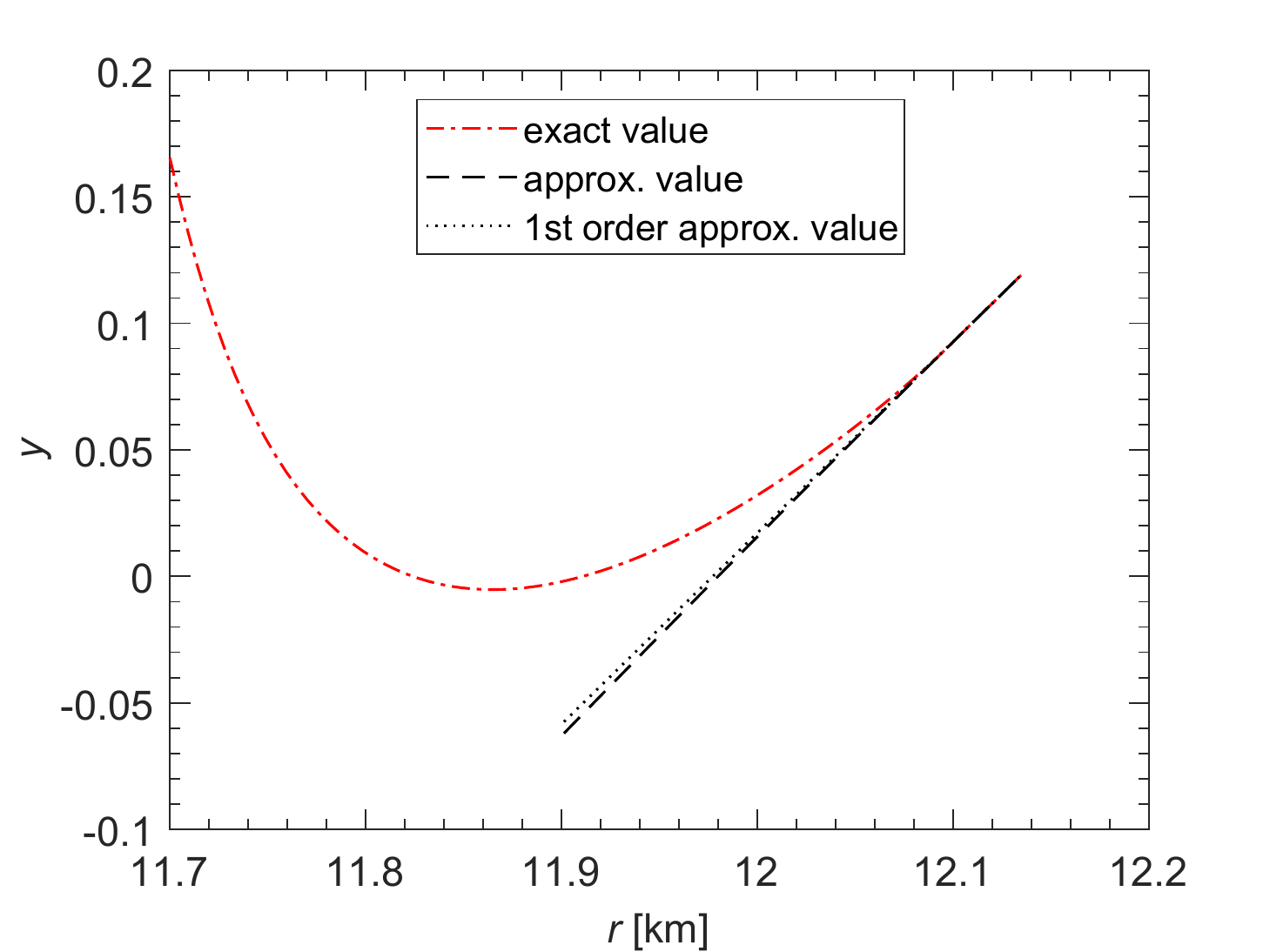}
\caption{(Color online) Variation of the function $y(r)$ with the radial coordinate $r$ for a NS with a mass $M=1.4M_\odot$. Exact values were calculated by solving numerically Eq.~(\ref{y_eq}) using the EoS without crust. Approximate values were obtained using Eq.~(\ref{y_eq_app}) and the first-order expansion (\ref{y_eq_app1}).}
\label{fig:Y_approx_HOM}
\end{center}
\end{figure}

\begin{figure}[h!]
\begin{center}
\includegraphics[width=\textwidth]{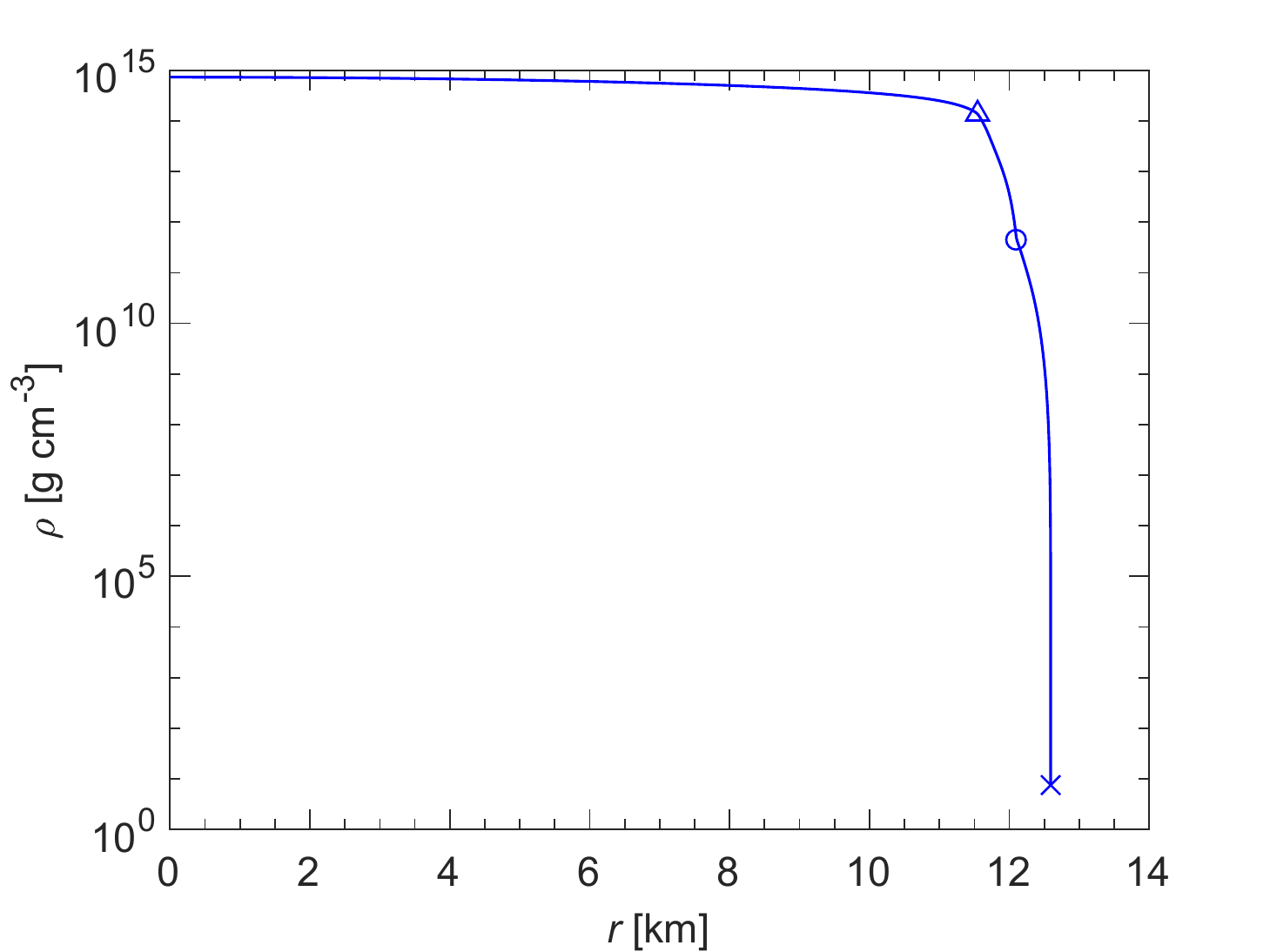}
\caption{Variation of the mass-energy density $\rho$ with the radial coordinate $r$ for a NS with crust having a  mass $M=1.4M_{\odot}$. The crust-core boundary and the neutron-drip transition (delimiting the inner and outer parts of the crust) are indicated by a triangle and a circle, respectively. The surface of the star is represented by a cross.}
\label{fig:RHO(R)}
\end{center}
\end{figure}

 As suggested in Figs.~\ref{fig:F(r)} and \ref{fig:Q(r)},
 \begin{equation}\label{eq:F_app}
    F(r) \approx \left(1-\frac{2 GM}{R c^2}\right)^{-1}\, , \hskip0.5cm Q(r)\approx -6F -4 C^2 F^2\, ,
\end{equation}
near the stellar surface. With these approximations, Eq.~(\ref{y_eq}) can be solved analytically: 
\begin{equation}\label{y_eq_app}
    y(r)\approx \frac{1}{2}\Biggl[-F+\sqrt{F^2-4Q}\tanh\left(\frac{1}{2}\sqrt{F^2-4Q}\ln\frac{r}{R}+\tanh^{-1}\frac{2 y_R+F}{\sqrt{F^2-4Q}}\right)\Biggr]\, . 
\end{equation}
To first order in $\epsilon=(r-R)/R$, the above expression reduces to 
\begin{equation}\label{y_eq_app1}
    y(r)\approx y_R - \left(\frac{r}{R}-1\right) (Q + F y_R + y_R^2)\, .
\end{equation}
As shown in Figs.~\ref{fig:Y_approx_CRUST} and \ref{fig:Y_approx_HOM}, Eq.~(\ref{y_eq_app}) and to a lesser extent its first-order approximation~\eqref{y_eq_app1}, are in very good agreement with numerical results, both for the EoSs with and without crust. This shows that the function $y(r)$ is 
essentially independent of the EoS for $r_\textrm{nd}\lesssim r\lesssim R$; however, the actual value of $R$ (hence also indirectly that of $y(R)$) is very sensitive to the EoS, as can be seen in Fig.~\ref{fig:M-R}. The EoS plays a direct role in the deeper region of the star, where the terms containing the energy density and the pressure in $F(r)$ and $Q(r)$ can no longer be ignored. With increasing depth, the term in the function $Q(r)$ containing $\mathcal{E}(r)c^2/c_s(r)^2$ indeed  rises sharply until the crust-core boundary is reached at the radial coordinate $r_\textrm{cc}$, reflecting the dramatic increase of the density $\rho=\mathcal{E}/c^2$ by about 14 orders of magnitude (see Fig.~\ref{fig:RHO(R)}) whereas the variation of the sound speed $c_s$ is comparatively small. Beneath the crust, the density increases only slightly while the sound speed increases substantially leading to a drop of the function $Q$. 
In the core, both the density and the sound speed are roughly constant so that the ratio $\mathcal{E}(r)c^2/c_s(r)^2$ remains small; the function $Q(r)$ thus exhibits a plateau close to $Q\approx Q(0)=-6$ since the terms containing the mass function $m(r)$ become also vanishing small as $r$ decreases. As shown in Fig.~\ref{fig:F(r)}, $F(r)\approx 1$ in the central core. The function $y(r)$ thus varies approximately as in Eq.~(\ref{y_eq_app}) with $F\approx 1$ and $Q\approx -6$:
\begin{equation}
    y(r)\approx \frac{1}{2}\Biggl[-1+5\tanh\left(\frac{5}{2}\ln\frac{r}{r_0}+\tanh^{-1}\frac{2 y(r_0)+F}{5}\right)\Biggr]\, ,
\end{equation}
where $r_0$ is a reference radial coordinate. Taking the limit $r_0\rightarrow 0$ with $y(r_0)=2$ thus leads to $y(r)\approx 2$: the function $y(r)$ is therefore approximately constant in the central core of the NS, as can be seen e.g. in Fig.~2 of Ref.~\cite{piekarewicz2019}.

\begin{figure}[ht]
\begin{center}
\includegraphics[width=\textwidth]{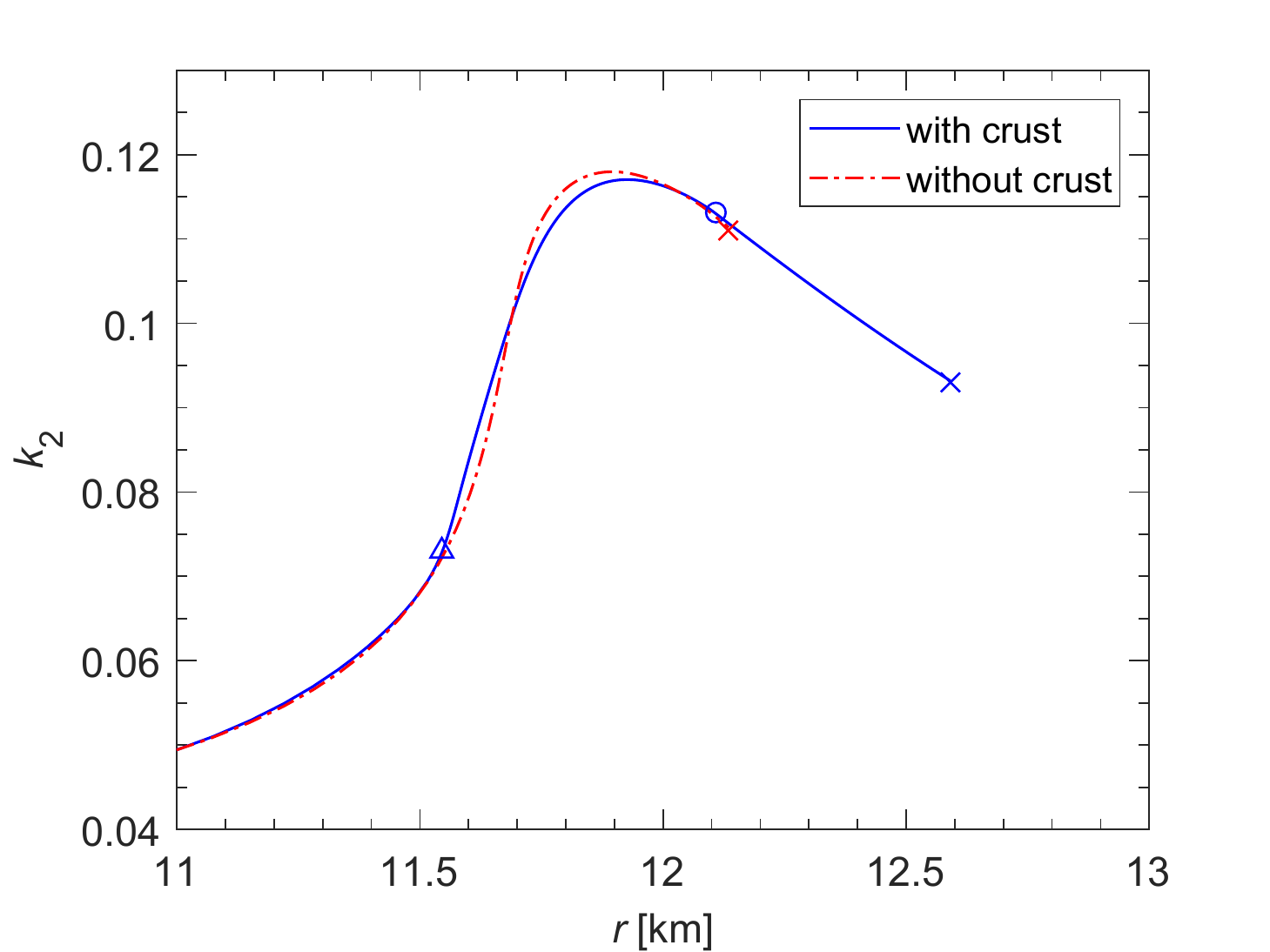}
\caption{(Color online) Local second gravito-electric Love number $k_2(r)$ as a function of the radial coordinate $r$ for a NS with and without crust, having a mass $M = 1.4M_\odot$. The crust-core boundary and the neutron-drip transition (delimiting the inner and outer parts of the crust) are indicated by a triangle and a circle, respectively. The surface of the star is represented by a cross. See text for details.}
\label{fig:k_2(r)}
\end{center}
\end{figure}

\begin{figure}[h!]
\begin{center}
\includegraphics[width=\textwidth]{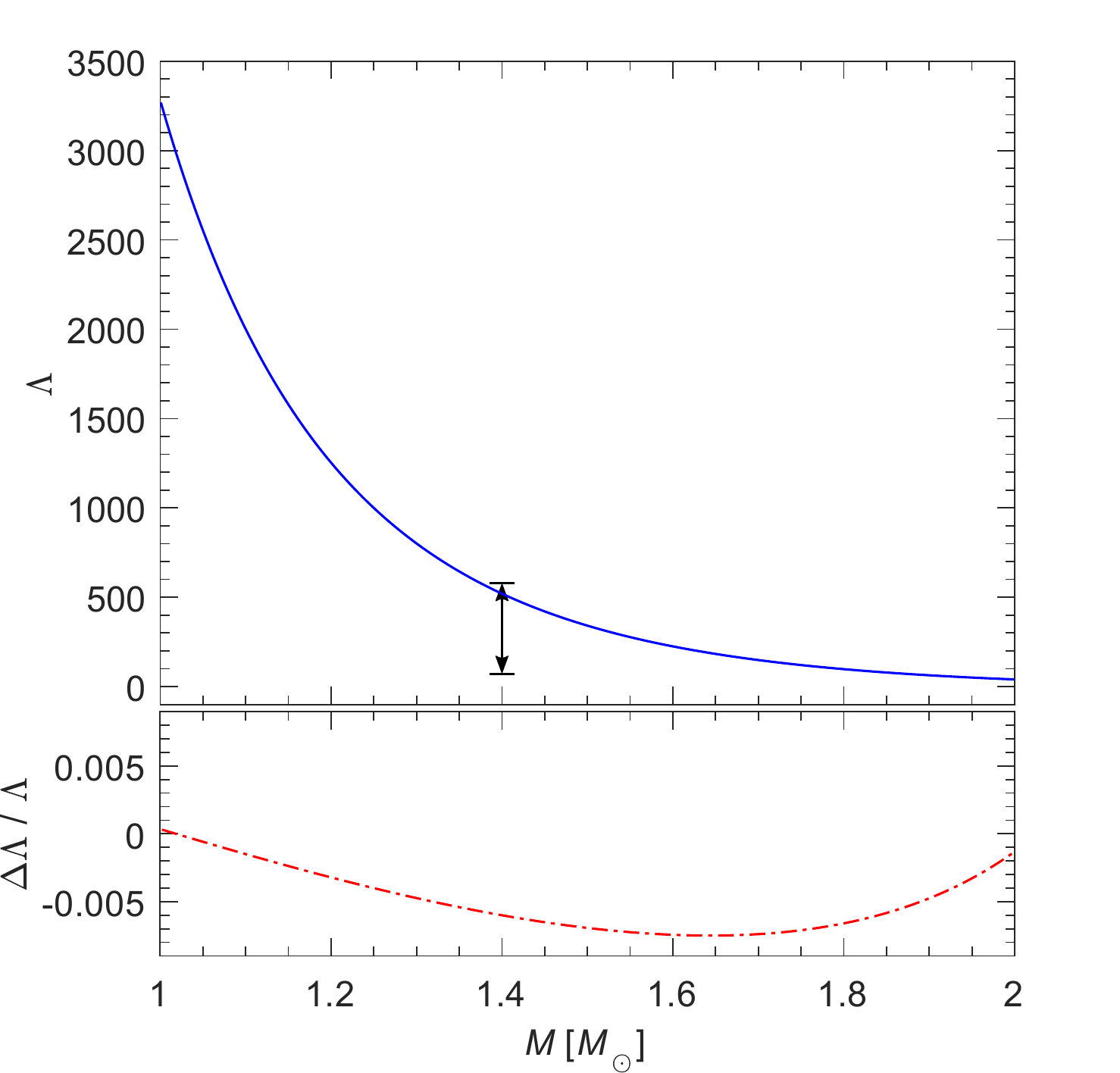}
\caption{(Color online) Top panel: dimensionless tidal deformability parameter $\Lambda$ as a function of NS mass $M$. Only the curve for the EoS with crust is shown, results for the EoS without crust being indistinguishable. The vertical double arrow represents the LIGO-Virgo constraint at 90\% confidence level from the analysis of GW170817~\cite{ligo2018}. Bottom panel: relative deviation $(\Lambda_{\rm hom}-\Lambda_{\rm crust})/\Lambda_{\rm crust}$ between the tidal deformability $\Lambda_{\rm crust}$ calculated with crust and the tidal deformability $\Lambda_{\rm hom}$ calculated without (homogeneous matter), as a function of NS mass $M$. See text for details.}
\label{fig:Lambda}
\end{center}
\end{figure}

The Love number $k_2$ obtained from the value of the function $y(r)$ at the stellar surface is very sensitive to the EoS but only through the stellar radius $R$. This is best seen in Fig.~\ref{fig:k_2(r)}, where we plotted the local Love number $k_2(r)$  obtained from Eq.~(\ref{eq:k2}) after replacing $C$ by $G m(r)/(r c^2)$ and $y_R$ by $y(r)$. Although the crust leads to a strong reduction of $k_2=k_2(R)$, its effect on the tidal deformability $\Lambda$ is mitigated by the increase in the radius $R$, as visible in Fig.~\ref{fig:Lambda} and previously noticed in Refs.~\cite{kalaitzis2019, piekarewicz2019,ji2019}. The impact of the crust on $\Lambda$ is found to be  significantly weaker than that previously reported in Ref.~\cite{kalaitzis2019}, the deviations amounting to 0.6\% for a $1.4M_\odot$ NS. Also represented in Fig.~\ref{fig:Lambda} is the LIGO-Virgo estimate $\Lambda_{1.4} = 190^{+390}_{-120}$ for the tidal deformability of a $1.4M_\odot$ NS at 90\% confidence level, as inferred from the analysis of  GW170817~\cite{ligo2018}. As previously shown in Ref.~\cite{perot2019}, the unified EoS based on BSk24 is consistent with this constraint.

\subsection{Love number and tidal deformability of realistic neutron stars from homogeneous matter}

Our analysis shows that the Love number of a realistic NS of mass $M$ and radius $R$ can be accurately estimated using the EoS of homogeneous matter only. It is enough to calculate the radius $R_0$ and the value $y_0\equiv y(R_0)$ of a NS without crust having the same mass $M$. From the formula obtained in Ref.~\cite{zdunik2017}, the radii $R$ and $R_0$ are related by 
\begin{equation}\label{eq:Rapp}
    R=R_0 \biggl\{ 1- \biggl[\left(\frac{m_\textrm{H}}{m_\textrm{Fe}}\right)^2-1\biggr]\left(\frac{R_0}{R_s}-1\right)\biggr\}^{-1}\, , 
\end{equation}
where $m_\textrm{Fe}$ denotes the atomic mass per nucleon of $^{56}$Fe, $m_\textrm{H}$ is the hydrogen mass (see Sec.~\ref{sec:unified_without}), and $R_s=2G M/c^2$ is the Schwarzschild radius. 
The value $y(R)$ can then be inferred from $y_0$ and $R_0$ as follows:  
\begin{equation}\label{y_eq_app0}
    y(R)\approx \frac{1}{2}\Biggl[-F_0+\sqrt{F_0^2-4Q_0}\tanh\left(\frac{1}{2}\sqrt{F_0^2-4Q_0}\ln\frac{R}{R_0}+\tanh^{-1}\frac{2 y_0+F_0}{\sqrt{F_0^2-4Q_0}}\right)\Biggr]
\end{equation}
with  
 \begin{equation}
    F_0 = \left(1-\frac{R_s}{R_0}\right)^{-1}\, , \hskip0.5cm Q_0=-6F_0-F_0^2 \left(\frac{R_s}{R_0}\right)^2\, .
\end{equation}
From the value $y(R)$, the Love number $k_2$ and the tidal deformability parameter $\Lambda$ can be readily calculated. Using $m_{\rm Fe}=930.412$~MeV and $m_{\rm H}=938.783$~MeV from Ref.~\cite{AME2016}, the relative errors on the radius $R$ and $y(R)$ for a $1.4 M_\odot$ NS amount to -0.13\% and 0.086\%, respectively. The corresponding errors on the Love number $k_2$ and on the tidal deformability parameter $\Lambda$ are respectively about -0.17\% and -0.82\%.

\section{Conclusion}
\label{sec:conclusion}

The role of the crust on the tidal deformability of a cold nonaccreted NS has been studied using the recent unified EoS BSk24~\cite{pearson2018}. This EoS, which was calculated in the framework of the nuclear-energy density functional theory, provides a thermodynamically consistent description of both the crust and the core of the star. The underlying functional was precision fitted to a large set of experimental and theoretical nuclear data~\cite{gcp2013}, including essentially all measured atomic masses as well as many-body calculations of pure neutron matter using realistic nucleon-nucleon potentials. The adiabatic index is found to vary substantially throughout the stellar interior, and therefore realistic EoSs can hardly be parametrized by simple polytropes. 

To better assess the influence of the crustal region on the global structure and on the tidal deformability of a NS, we have calculated the EoS of a putative NS made entirely of homogeneous matter using the same functional. The presence of the crust leads to NSs with larger radii, however the differences remain small for realistic stellar masses (the radius is increased by 6\% at most for a one-solar mass star). On the contrary, the crust plays a more significant role in the determination of the Love number $k_2$, especially for low-mass stars. NSs with crust are thus found to have smaller Love numbers than purely homogeneous stars. This reduction mainly arises from the increase in the stellar radius but is essentially insensitive to the details of the low-density part of the EoS. Although the crust leads to a strong reduction of $k_2$, its effect on the tidal deformability parameter $\Lambda$ is mitigated by the increase in the radius $R$, as previously noticed in Refs.~\cite{kalaitzis2019, piekarewicz2019}. However, contrary to Ref.~\cite{kalaitzis2019}, the deviations we found never exceed 0.75\%. Our analysis of the tidal-deformability equations allows for a simple and very accurate analytic estimate of the Love number $k_2$ and tidal deformability $\Lambda$ of realistic NSs using the EoS of homogeneous matter only, through Eqs.~(\ref{eq:k2}), (\ref{eq:Rapp}), and (\ref{y_eq_app0}). 

In our calculations, we did not account for the possible existence of a mantle of nuclear pastas at the interface between the crust and the core regions. However, its effect on the Love number and tidal deformability is expected to be negligible since the deviations in the EoS are found to lie within the fitting errors~\cite{pearson2019}. We have also ignored the role of elasticity. Although the corrections to the Love number are expected to be much smaller than the deviations found here~\cite{penner2012,biswas2019}, they may still be of relevance for the third generation of gravitational-wave detectors. Finally, the presence of a neutron superfluid in the inner crust also requires further scrutiny.

\section*{Acknowledgments}
The authors thank Leszek Zdunik and Morgane Fortin for valuable discussions. This work was financially supported by the Fonds de la Recherche Scientifique - FNRS (Belgium) under Grants No. CDR J.0115.18. and no. 1.B.410.18F, and the European Cooperation in Science and Technology Action (EU) CA16214. L. Perot is a FRIA grantee of the Fonds de la Recherche Scientifique - FNRS  (Belgium).

\bibliography{biblio}
\end{document}